\documentclass[conference]{IEEEtran}
\IEEEoverridecommandlockouts
\usepackage{cite}
\usepackage{amsmath,amssymb,amsfonts,bm}
\usepackage{paralist}
\usepackage{float}
\usepackage{algorithmic}
\usepackage{graphicx}
\usepackage{textcomp}
\usepackage{xcolor}
\def\BibTeX{{\rm B\kern-.05em{\sc i\kern-.025em b}\kern-.08em
    T\kern-.1667em\lower.7ex\hbox{E}\kern-.125emX}}
\usepackage{soul}
\usepackage[skip=0pt,font=smaller]{subcaption}
\usepackage{array}

\begin{document}

\title{Accelerating Relational Database Analytical Processing with Bulk-Bitwise Processing-in-Memory
\thanks{This work was supported by the European Research Council through the European Union's Horizon 2020 Research and Innovation Programme under Grant 757259 and through the European Union's Horizon Europe Research and Innovation Programme under Grant 101069336.}
}

\author{\IEEEauthorblockN{Ben Perach \qquad Ronny Ronen \qquad Shahar Kvatinsky}
\IEEEauthorblockA{\textit{Viterbi Faculty of Electrical \& Computer Engineering} \\
\textit{Technion -- Israel Institute of Technology}\\
Haifa, Israel \\
benperach@campus.technion.ac.il \quad ronny.ronen@ef.technion.ac.il \quad shahar@ee.technion.ac.il}\vspace{-18pt}
}

\maketitle

\begin{abstract}
Online Analytical Processing (OLAP) for relational databases is a business decision support application. The application receives queries about the business database, usually requesting to summarize many database records, and produces few results. Existing OLAP requires transferring a large amount of data between the memory and the CPU, having a few operations per datum, and producing a small output. Hence, OLAP is a good candidate for processing-in-memory (PIM), where computation is performed where the data is stored, thus accelerating applications by reducing data movement between the memory and CPU. In particular, bulk-bitwise PIM, where the memory array is a bit-vector processing unit, seems a good match for OLAP. With the extensive inherent parallelism and minimal data movement of bulk-bitwise PIM, OLAP applications can process the entire database in parallel in memory, transferring only the results to the CPU. This paper shows a full stack adaptation of a bulk-bitwise PIM, from compiling SQL to hardware implementation, for supporting OLAP applications.
Evaluating the Star Schema Benchmark (\textit{SSB}), bulk-bitwise PIM achieves a $4.65\times$ speedup over Monet-DB, a standard database system.
\end{abstract}

\vspace{-15pt}
\section{Introduction}

Analytical processing of relational databases is a business decision support application, allowing decision-makers to analyze their business data~\cite{Kimball2013}. Often, these analyses require summarizing large sections of the database, taking a long execution time~\cite{Hasso2013}, mostly on transferring data from the memory to the CPU. Summarizing large sets of data to few results hints that this application can benefit from \textit{processing-in-memory (PIM)} techniques.

PIM techniques come in various types and technologies~\cite{PIMDB,AMBIT,SIMDRAM,MOUSE,Pinatubo,ROWCLONE,AquaboltXL,Reuben2017}, all operate on data at (or close to) where it is stored, \textit{i.e.}, the memory. In this paper, we are interested in accelerating analytical processing of relational databases with a specific PIM technique called \textit{bulk-bitwise PIM}~\cite{AMBIT,SIMDRAM,Pinatubo,PIMDB,MOUSE}. Bulk-bitwise PIM is characterized by utilizing the memory cell arrays to both process the data and directly store the result. Thus, computing with bulk-bitwise PIM can minimize data movement. As data-intensive applications invest most of their time and energy in memory access~\cite{Wulf1995,Johannes2018,DarkMemory}, our approach for accelerating analytical processing consists of reducing the required memory accesses by the host. We do so by leveraging the in-place processing of bulk-bitwise PIM to distill the information stored in memory, transferring as little information as possible to the host. Data transfer reduction is possible because databases can be transferred into PIM memory once and used many times.

Note that using bulk-bitwise PIM in this way is orthogonal to the host processor type (\textit{e.g.}, CPU, GPU, FPGA) since any host requires to get the information from the memory. Furthermore, using bulk-bitwise PIM to reduce the data movement out of the memory cell arrays is complementary to other PIM techniques where dedicated processing units are placed close to the memory arrays. These dedicated processing units can act as the hosts for the bulk-bitwise PIM technique, profiting from the same reduced data transfer.

In this paper, we combine the thread from several of our previous works~\cite{PIMDB,PIMConsistency,DACpaper} to show how analytical processing for relational databases can be supported with bulk-bitwise PIM. As the processing capabilities of bulk-bitwise PIM depend on the data arrangement within the memory arrays, we show how to arrange the database relations within the memory arrays, creating a dedicated data structure for bulk-bitwise PIM. We then identify the basic primitive of bulk-bitwise PIM that can reduce data transfer with this new data structure (\textit{i.e.}, filter and aggregate). Afterward, we show how more complex database operations are supported for the \textit{star schema} database~\cite{SSB,Kimball2013} (\textit{i.e.}, GROUP-BY and JOIN), allowing execution of full queries. Using this support, we can execute the full Star Schema Benchmark (SSB)~\cite{SSB}. We evaluate these techniques with a memristive bulk-bitwise PIM design~\cite{PIMDB,PIMConsistency} using the gem5~\cite{gem5} simulation environment.
\vspace{-4 pt}
\section{Background}
\vspace{-2 pt}
\subsection{Relational Databases and Analytical Processing}

\begin{figure}[!t]
\centering
\begin{minipage}[c]{0.05\columnwidth}
\begin{subfigure}[c]{\textwidth}
\caption{}\label{subfig:relation}
\end{subfigure}
\end{minipage}%
\begin{minipage}[c]{0.48\columnwidth}
\includegraphics[width=\textwidth]{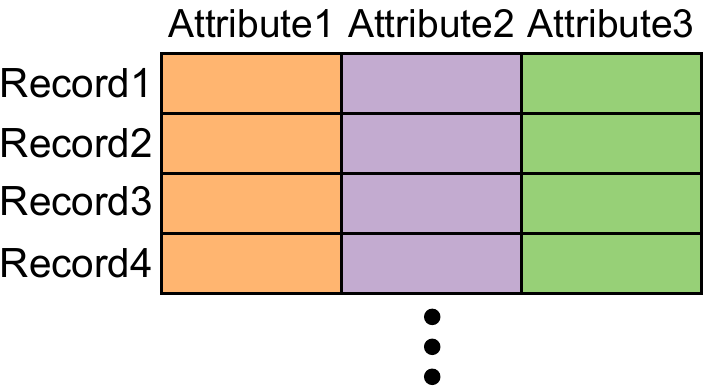}
\end{minipage}%
\begin{minipage}[c]{0.05\columnwidth}
\begin{subfigure}[c]{\textwidth}
\caption{}\label{subfig:relation_xb}
\end{subfigure}
\end{minipage}%
\begin{minipage}[c]{0.42\columnwidth}
\includegraphics[width=\textwidth]{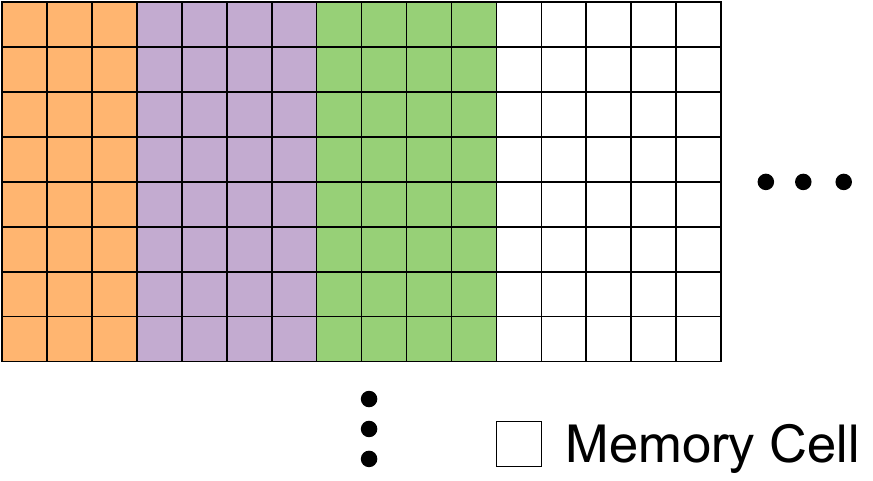}
\end{minipage}%
\vspace{-5 pt}
\caption{(a) A database relation. (b) The embedding data structure within a memory cell array.}
\label{fig:relation}
\vspace{-22 pt}
\end{figure}

The relational database is a data model organized into one or more relations (tables). Each relation is constructed as multiple records and attributes (shown in Fig.~\ref{subfig:relation}), represented by the relation's rows and columns, respectively. Records are independent, holding information belonging to a single item with a single value for each attribute. Each relation has an attribute (or a set of attributes) that uniquely identifies the records in the relation and is called the \textit{key} of that relation. When a relation has an attribute that has values from a key of another relation, this attribute is called a \textit{foreign key}.

Queries on the database are questions about the data held in the database. In analytical processing, queries require finding all records fulfilling certain conditions on their attributes and summarizing (\textit{e.g.}, sum, average, max) one or more attributes across the found records~\cite{Hasso2013,Kimball2013}. This summarizing can also be requested per subgroups of the found records, where subgroups are defined according to unique values of some attributes. This division for subgroups is called a \textit{GROUP-BY} operation.  

When a query includes conditions involving attributes from multiple relations, records from the different relations are matched according to these conditions. The operation of matching records is called a \textit{JOIN}. When the condition between the relations' attributes for JOIN is equality, the JOIN is named \textit{equi-JOIN}. For analytical processing queries, JOIN operations are usually the most time-consuming part of query execution~\cite{Dreseler2020}. 

\begin{figure}[!t]
\centering
\includegraphics[width=\columnwidth]{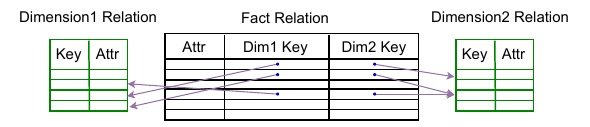}%
\caption{A star schema illustration.}
\label{fig:star_schema}
\vspace{-22 pt}
\end{figure}

A common database structure for analytical processing is the \textit{star schema}~\cite{Kimball2013,SSB}. An illustration of the star schema structure is shown in Fig.~\ref{fig:star_schema}. In this schema, there is a single large relation, called the \textit{fact} relation, and several small relations called the \textit{dimension} relations. The fact relation has a foreign key to each of the dimension relations. In the star schema, JOIN operations required by queries are, by and large, only equi-JOIN between a dimension relation's key and its respective fact relation foreign key~\cite{Kimball2013}. All the SSB benchmark~\cite{SSB} queries use only this kind of JOIN.

\subsection{Bulk-Bitwise PIM}

\begin{figure}[!t]
\centering
\includegraphics[width=0.60\columnwidth]{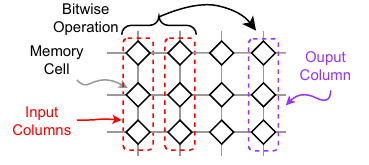}%
\vspace{-7 pt}
\caption{A $3\times4$ memory array with a column-wise logic operation.}
\label{fig:bulkbitwisePIM}
\vspace{-14 pt}
\end{figure}

Bulk-bitwise PIM uses the memory cell arrays as processing units, which can be implemented with DRAM or emerging nonvolatile memory technologies~\cite{AMBIT,SIMDRAM,Pinatubo,PIMDB,MOUSE}. Because of the regular structure of these arrays, the supported operations are bitwise logic operations (\textit{e.g.}, AND, NOT, NOR) between array rows or columns (shown in Fig.~\ref{fig:bulkbitwisePIM}). When multiple memory cell arrays operate concurrently, the effective operation is a very wide bitwise operation, \textit{i.e.}, bulk-bitwise operations. As such, bulk-bitwise PIM can process data where it is stored and exhibit high computational bandwidth.

To support virtual memory, bulk-bitwise PIM operations are restricted to use and rewrite data within a single virtual page~\cite{PIMDB}, usually a huge page. This way, when a program sends a PIM operation to memory with a virtual address, the virtual address can be translated in the standard fashion and the PIM operation is routed by the hardware to its designated place. To perform the same operation on several pages, however, the same operation has to be sent to each page separately. In addition, since the PIM computation is tightly coupled with the layout of data in the memory cell arrays, data for PIM has to be structured in a specific, dedicated way. To allow software in virtual memory fine-grain control over the layout of data structures within the cell arrays, the mapping of addresses to cell array location is part of the bulk-bitwise programming model~\cite{PIMDB}. The specified mapping is on the page offset bits of the virtual address since they do not change on virtual-physical address translation, giving the virtual space software control over these bits. 

To guarantee program correctness, the ordering rules of PIM operations with the standard memory operations (\textit{e.g.}, loads, stores) have to be defined~\cite{PIMConsistency}. Having well-defined ordering rules for bulk-bitwise PIM requires the PIM memory and host caches to be coherent, supported by the host hardware~\cite{PIMConsistency}.

\vspace{-4pt}
\section{The Relation Data Structure}
\vspace{-4pt}
A database relation is stored in a memory cell array, as shown in Fig.~\ref{fig:relation}. It is necessary to store data in a certain way so it can be directly processed without further movement since the processing capabilities of bulk-bitwise PIM are directly connected to the layout of data within the memory cell array.
To this end, each record is set across a cell array row, and each attribute spans several columns.
In this data structure, column-wise operations can be used to process all records in parallel. Some cell array columns, however, have to remain empty to store the PIM results before they are read from the memory. Note that a record's bytes are not consecutive in virtual address space due to the address mapping.

If a single array row is not enough to hold all the attributes of a relation, the relation's attributes must be split on more than a single cell array. These additional cell arrays are put on different pages and might require moving data between the pages using standard loads and stores.

\vspace{-2 pt}
\section{Bulk-Bitwise PIM Primitives}
\vspace{-2 pt}
This section presents the basic database operations using the relation data structure and bulk-bitwise PIM capabilities. These database primitives reduce the required data transfer between the host and memory.

\subsection{Filter}
\label{subsec:filter}
\begin{figure}[!t]
\centering
\begin{minipage}[c]{0.05\columnwidth}
\begin{subfigure}[c]{\textwidth}
\caption{}\label{subfig:filter}
\end{subfigure}
\end{minipage}%
\begin{minipage}[c]{0.4\columnwidth}
\includegraphics[width=\textwidth]{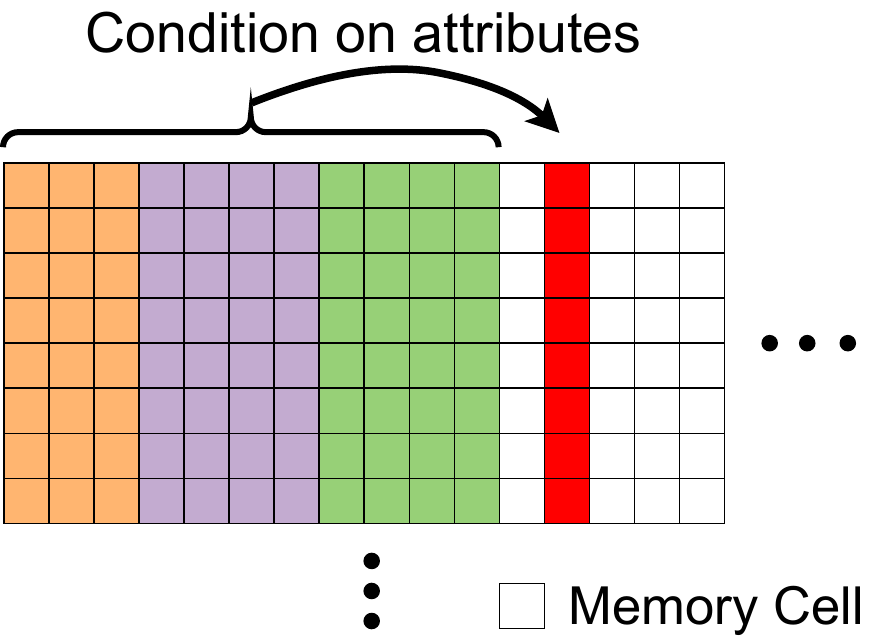}
\end{minipage}%
\begin{minipage}[c]{0.05\columnwidth}
\begin{subfigure}[c]{\textwidth}
\caption{}\label{subfig:aggregation}
\end{subfigure}
\end{minipage}%
\begin{minipage}[c]{0.4\columnwidth}
\includegraphics[width=\textwidth]{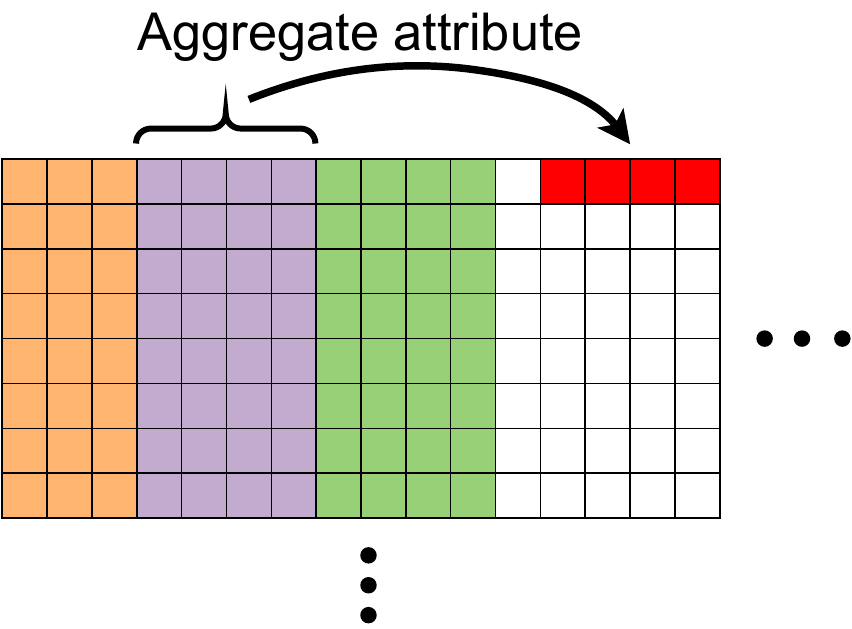}
\end{minipage}%
\caption{Illustration of PIM primitives: (a) filter and (b) aggregation. The results are marked with red cells.}
\label{fig:primitives}
\vspace{-18 pt}
\end{figure}

A filter operation, shown in Fig.~\ref{subfig:filter}, checks a condition across all relation records. This operation filters all records for a query or a single subgroup in a GROUP-BY operation (Section~\ref{subsec:groupby}). The condition checking is done with PIM,  where the result is a single bit per record. The resulting bit is stored in the same cell array column across all cell arrays of the relation. Hence, to assert which record passed the condition, the host only needs to read a single bit per record instead of the condition's attributes per record. The data transfer reduction depends on the number of attributes in the condition, the attributes' lengths, and the data itself (as non-PIM techniques are data depended~\cite{Hasso2013}).

To support the filter primitive, the PIM module instruction set includes comparison operations (\textit{e.g.}, equality, less-than), logic operations (\textit{e.g.}, AND, OR, NOT), and arithmetic operations (\textit{e.g.}, addition, multiplication). These operations are supported for both an attribute with an attribute and an attribute with an immediate. Additionally, these operations are supported for multiple attribute lengths.

\vspace{-2pt}
\subsection{Aggregation}

The aggregation operation, shown in Fig.~\ref{subfig:aggregation}, reduces a specific relation attribute, across all of the relations, to a single value (\textit{e.g.}, sum, average, max). This reduction, however, includes values only from selected records. 
This selection is done by first filtering the records according to the condition (as in the filter primitive in Section~\ref{subsec:filter}). Instead of reading the filter result, it is used to generate a masked version of the attribute to be aggregated (without overwriting the original attribute). The mask operation nullifies the non-selected values so they will not affect the aggregation operation. 
The masked attribute is then aggregated, according to the desired operation, within each cell array, resulting in a single value per cell array. Afterward, the host reads the single value from each cell array and aggregates them to complete the operation. Hence, the host only has to read a single value per cell array per aggregation.

For example, when summing an attribute, the attribute's bits are ANDed with the filter result bit to create the masked attribute. Hence, the unselected records have a zero value in the masked attribute, while the selected records' values remain the same. Summing the masked attribute will have the same result as summing only the attribute for the selected records. The host reads the single values in each cell array and sums them together, producing the final result.
 
Since the aggregation operations are performed value by value, they must be commutative and associative. Hence, the PIM module supports the aggregation operation for sum, min, and max within each cell array. Other aggregation operations can be supported as a combination of commutative and associative aggregation. For example, to perform an average (a non-associative operation) on an attribute, the attribute is first summed. Then, the filter result is also aggregated using sum, resulting in the count of the selected records. The host then divides the total sum of the records by their count to produce the average.

\subsection{Evaluating PIM Primitives}

To evaluate the performance of the PIM primitives, we used the TPC-H benchmark~\cite{TPCH}, a relation database analytical processing benchmark. The baseline compared to is an in-house implementation of an in-memory database~\cite{Hasso2013}, where the entire database is stored in DRAM main memory. The baseline ran on the same system as with PIM and executed the same query section as the PIM system. A full evaluation description and more analysis are presented in~\cite{PIMDB}.

\begin{figure}[!t]
\centering
\begin{minipage}[c]{0.05\columnwidth}
\begin{subfigure}[c]{\textwidth}
\caption{}\label{subfig:filter_only}
\end{subfigure}
\end{minipage}%
\begin{minipage}[c]{0.64\columnwidth}
\includegraphics[width=\textwidth]{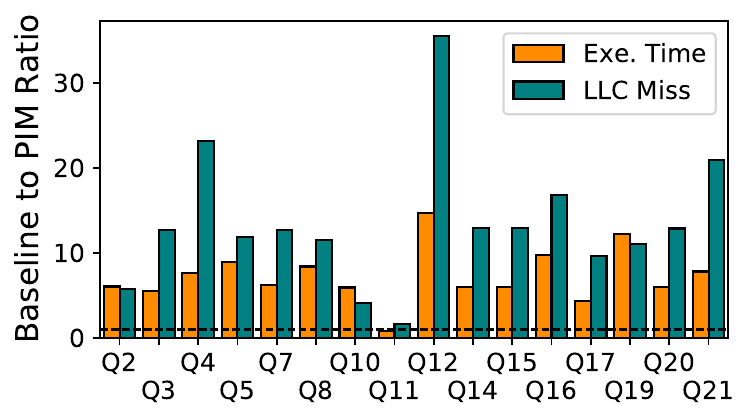}
\end{minipage}%
\begin{minipage}[c]{0.05\columnwidth}
\begin{subfigure}[c]{\textwidth}
\caption{}\label{subfig:full_queries}
\end{subfigure}
\end{minipage}%
\begin{minipage}[c]{0.27\columnwidth}
\includegraphics[width=\textwidth]{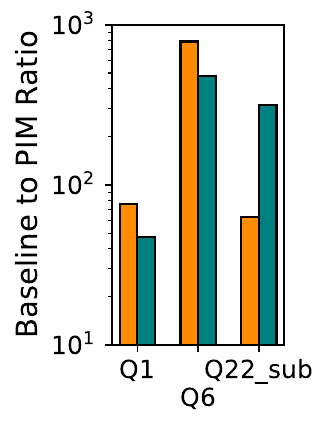}
\end{minipage}%
\vspace{-9pt}
\caption{Primitive operation speedup and data transfer reduction (last-level cache misses) on the TPC-H benchmark. (a) Filter-only and (b) full queries.}
\label{fig:runtime}
\vspace{-19 pt}
\end{figure}

The speedup and memory accesses reduction (LLC misses) for the PIM over the baseline are shown in Fig.~\ref{fig:runtime}. The queries are divided into two groups, \textit{full} queries and \textit{filter-only} queries. Full queries can perform the entire query, including the required aggregations, while filter-only queries can perform only the filter part of the query. Since the aggregation performs a more substantial data transfer reduction, it also achieves a significantly higher speedup. We also see that the data transfer reduction is similar in magnitude to the speedup, supporting our approach that bulk-bitwise PIM speedup comes from the data transfer reduction.

\vspace{-4pt}
\section{Supporting the Star Schema}

Using the PIM database primitives and understanding their strengths and weaknesses, more complex operations can be designed and supported. This section presents the support for JOIN and GROUP-BY for the star schema. Full details and full evaluation are presented in~\cite{DACpaper}.

\subsection{Supporting JOIN}

JOIN requires matching records from different relations, where the matching itself is data-dependent. Hence, performing JOIN requires many data movements, which conflicts with the goal of data movement reduction. To avoid this, JOIN is accomplished by pre-computing the required JOIN output and storing the relations as a single JOINed relation. Pre-computing a JOIN operation is a known technique to accelerate query execution, appearing as denormalization~\cite{Shin2006} or materialized views~\cite{Chirkova2012}. Storing a pre-computed JOIN, however, has drawbacks. Characteristics of the star schema and bulk-bitwise PIM can mitigate these drawbacks.

\begin{figure*}[t!]
\centering
\includegraphics[width=0.95\textwidth]{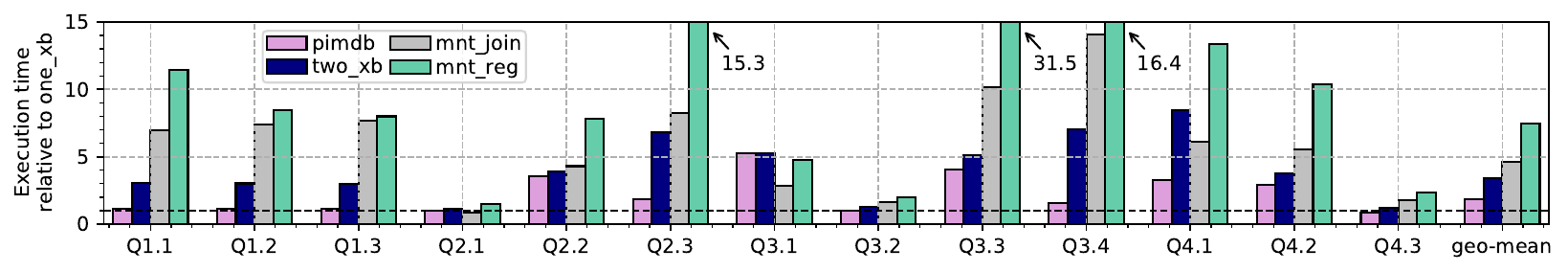} 
\vspace{-6pt}
\caption{Execution time relative to one\_xb for the Star Schema Benchmark (SSB) benchmark queries.}
\vspace{-20pt}
\label{fig:latency}
\end{figure*}

A significant drawback of pre-computed JOINs is their flexibility. If a query requires a different JOIN than the one pre-computed, then the pre-computation is not helpful, and the required JOIN has to be performed. For the star schema, most JOIN operations are an equi-JOIN between a dimension relation and the fact relation on the dimension key~\cite{Kimball2013,SSB}. All of the JOIN operations of the SSB~\cite{SSB} benchmark are of this type. Hence, we cover and accelerate the vast majority of cases by storing the fact relation equi-JOINed to all dimension relations on their key.

Using pre-computed JOINs also complicates maintenance~\cite{Shin2006,Chirkova2012}. Since JOIN operations duplicate data, performing an UPDATE on a pre-computed JOIN requires changing many entries. Using bulk-bitwise PIM, however, alleviates this drawback. The filter primitive can efficiently find all records to update. Furthermore, PIM operations can perform the update itself, not requiring reading any value from PIM memory (\textit{i.e.}, implementing a PIM MUX with the filter result as select~\cite{DACpaper}). 

\vspace{-2pt}
\subsection{Supporting GROUP-BY}
\label{subsec:groupby}

Performing GROUP-BY can be simply done by using PIM aggregation for every subgroup, aggregating subgroups one by one. The number of subgroups, however, can be large. Performing many aggregations may result in long latency, high energy, and high power, and reduce the lifetime of the PIM module due to the limited endurance of memory cells~\cite{PIMDB}.

To mitigate these deficiencies of PIM aggregation, a standard CMOS logic circuit is added to the memory cell array peripherals. The circuit performs the aggregation by reading the masked attribute to be aggregate value by value from its cell array. The circuit aggregates the read values and writes the final result back to the cell array. Thus, the latency, energy, and power of the aggregation primitive are reduced compared to a pure bulk-bitwise PIM aggregation. Furthermore, the PIM module lifetime is increased, as cells are written only at the end of the aggregation operation.

Nevertheless, the latency of the entire GROUP-BY operation can still be high if there are many subgroups. We note that aggregation can be performed in another way. The records for the query (for all subgroups) can be filtered, then read record-by-record by the host. The host then classifies and aggregates the records in their subgroups. This host aggregation technique is dependent on the number of records in the query and is independent of the number of subgroups. PIM aggregation, however, depends on the number of subgroups and not the number of records in the query. Hence, it is beneficial to have PIM aggregating a few large subgroups, leaving many small subgroups to host aggregation, exploiting the strength of each method. To divide the subgroups between PIM and host aggregation, a performance model is used along with an estimate for subgroups' sizes~\cite{DACpaper}. This estimate is done by taking a small sample from the database. This GROUP-BY technique is adapted from an in-cloud processing work~\cite{PushdownDB}.

\vspace{-2pt}
\subsection{Star Schema Evaluation}

We evaluate our solutions on the SSB benchmark~\cite{SSB}. The execution times are shown in Fig.~\ref{fig:latency}. We compared three versions of PIM. \textit{one-xb} stores records of the JOINed relation in a single memory cell array. \textit{two-xb} is where the JOINed relation's records are vertically split into two memory cell arrays, having the fact relation's attributes in one array and all of the dimension relations' attributes in another array. \textit{pimdb} is the same as \textit{one-xb}, only the PIM aggregation (PIM-gb) is performed with pure bulk-bitwise operations (without the added aggregation circuit).
Two baselines are compared, \textit{mnt-reg} and \textit{mnt-join}. These baselines run Monet-DB~\cite{MonetDB}, a real in-memory database system running on real hardware. The \textit{mnt-reg} and \textit{mnt-join} hold the SSB database in its standard and pre-computed JOIN, respectively. 

The best execution time is for \textit{one-xb}, having a geo-mean speedup of $7.46\times$ and $4.65\times$ over \textit{mnt-reg} and \textit{mnt-join}, respectively. \textit{two-xb} is $3.39\times$ slower (in geo-mean) than \textit{one-xb} since many data transfers are required between the memory cell arrays of the relation. \textit{two-xb}, however, still has $1.37\times$ speedup on \textit{mnt-join}.
The cases where a non-PIM execution is faster than a PIM execution (Q2.1, Q3.1, and Q4.1) are where the PIM execution does not achieve a significant data transfer reduction. In these cases, the number of records required by the query is high. Due to the data structure of the PIM relation, most of the relation's records are transferred to the host, resulting in little to no data reduction and removing the advantage of PIM. See~\cite{DACpaper} for more details and discussions.


\section{Conclusions}

This paper shows how to support analytical processing of relational databases using bulk-bitwise PIM. Our bulk-bitwise PIM technique aimed to reduce the required data transfers. By substituting serial data accesses to memory with very wide and short operations within the memory, we achieve a significant speedup over von Neumann machines.

We first designed a data structure suited for bulk-bitwise PIM. Then, we identified and supported primitive operations (filter and aggregate), performing relevant functionality and reducing data transfer. These primitives were evaluated and studied. Based on these primitives, we support more complex operations (JOIN and GROUP-BY) and evaluated a full database benchmark (SSB) for a system based on memristive bulk-bitwise PIM. We believe this work will inspire other research for further adaptation of applications for bulk-bitwise PIM.

\bibliographystyle{IEEEtran}
\bibliography{IEEEabrv,ref}

\end{document}